\documentclass[english,smallextended]{article}       

\usepackage[pdftex]{graphicx}
\usepackage{amssymb}
\usepackage{amsmath}
\usepackage{color}
\usepackage{authblk}
\DeclareMathOperator{\Tr}{Tr}

\usepackage{mciteplus}
\usepackage[hidelinks,bookmarksopen]{hyperref}
\usepackage{epstopdf}
\begin{document}

\title{Bandwidth-limited and noisy pulse sequences for single qubit operations in semiconductor spin qubits}

\author{E. Ferraro}
\author{M. De Michielis}

\affil{CNR-IMM, Unit of Agrate Brianza, Via C. Olivetti 2, 20864 Agrate Brianza (MB), Italy}

\maketitle

\begin{abstract}
Spin qubits are very valuable and scalable candidates in the area of quantum computation and simulation applications. In the last decades, they have been deeply investigated from a theoretical point of view and realized on the scale of few devices in the laboratories. In semiconductors, spin qubits can be built confining the spin of electrons in electrostatically defined quantum dots. Through this approach it is possible to create different implementations: single electron spin qubit, singlet-triplet spin qubit, or a three electrons architecture, e. g. the hybrid qubit. For each qubit type, we study the single qubit rotations along the principal axis of Bloch sphere including the mandatory non-idealities of the control signals that realize the gate operations. The realistic transient of the control signal pulses are obtained by adopting an appropriate low-pass filter function. In addition the effect of disturbances on the input signals is taken into account by using a Gaussian noise model.
\end{abstract}


\section{Introduction}
The confinement of electron spins in host semiconductors is intensely studied in view of powerful applications in quantum computation and simulation \cite{Hanson-2007, Vandersypen-2017, Russ-2017,Rotta-2017}. Largely studied are qubit defined in electrostatically or self-assembled QDs \cite{Loss-1998, Veldhorst-2014, Kawakami-2014}, donor spins in solid matrices \cite{Kane-1998, Pla-2012, Pla-2013} or a combination of them \cite{Pica-2016, Harvey-2017}. In particular, semiconductor-based qubits assure long electron spin coherence times, easy manipulation, fast gate operations, and potential for scaling, in addition to the compatibility with the existing CMOS process. Starting from analytical expressions for the realization of logical gates, that are rotations along the main axis on the Bloch sphere, we studied how such operations are affected when non-idealities are included in the control pulses. The control and the manipulation of the qubit need electric voltages or magnetic fields, using DC pulses or microwave drives depending of the qubit type. We focus our study on three QD spin qubits: the single spin qubit \cite{Morton-2011}, the singlet-triplet qubit \cite{Levy-2002} and the hybrid qubit \cite{Shi-2012}.

The single spin qubit is realized confining the spin of a single electron in a single QD. The logical basis is defined adopting the two spin eigenstates $|\!\uparrow\rangle$ and $|\!\downarrow\rangle$ as the logical $|0\rangle$ and $|1\rangle$ respectively. In a rotating frame, which rotates at the angular frequency $\omega$, under the rotating wave approximation (RWA), the Hamiltonian is given by: 
\begin{equation}\label{HSS2}
H_{SS}=\frac{\hbar}{2}\Delta\omega_z\sigma^z + \frac{\hbar}{2}\Omega\cos(\phi)\sigma^x+\frac{\hbar}{2}\Omega\sin(\phi)\sigma^y, 
\end{equation}
where $\sigma^{z(x,y)}$ is the Pauli operator, $\Delta\omega_z\equiv\omega_z-\omega$ where $\omega_z$ is the Larmor angular frequency, $\hbar\omega_z=g_e\mu_BB_0$ is the Zeeman energy associated to the constant magnetic field $B_0$ applied in the $z$ direction ($g_e$ is the electron g-factor and $\mu_B$ is the Bohr magneton) and $\hbar\Omega=g_e\mu_BB_1/2$ is the Zeeman energy associated to the oscillating magnetic field $B_1$, with phase $\phi$ and angular frequency $\omega$. Qubit manipulation is obtained by modulating the phase $\phi$.  

The singlet-triplet qubit is defined on states of two electrons confined in a double QD. The logical states are a superposition of two-particle spin singlet and triplet states, that are $|0\rangle\equiv|S\rangle$ and $|1\rangle\equiv|T_0\rangle$, where each QD is occupied with one electron. The effective Hamiltonian model is
\begin{equation}\label{singlet-triplet}
H_{ST}=\frac{1}{2}\Delta E_z(\sigma_{1}^z-\sigma_{2}^z)+\frac{1}{4}J\boldsymbol{\sigma}_{1}\cdot\boldsymbol{\sigma}_{2},
\end{equation}
where $\boldsymbol{\sigma}_1$ and $\boldsymbol{\sigma}_2$ are the Pauli matrices referring to the two electrons. The first term on the right side is the Zeeman energy with $\Delta E_z\equiv\frac{1}{2}(E_1^z-E_2^z)$, namely a magnetic field gradient between the QDs, the second term is the exchange interaction through the coupling constant $J$. The ST qubit allows fast readout and fast manipulation as long as local magnetic gradient have been created through, for example, the use of a micro-magnet in close proximity.

The hybrid qubit is realized confining three electrons in a double QD. The logical states (coded in the $S=\frac{1}{2}$ and $S_z=\frac{1}{2}$ three electrons subspace) have been expressed by $|0\rangle\equiv|S\rangle|\!\uparrow\rangle$ and $|1\rangle\equiv\sqrt{\frac{1}{3}}|T_0\rangle|\!\uparrow\rangle-\sqrt{\frac{2}{3}}|T_+\rangle|\!\downarrow\rangle$, where combined singlet and triplet states of a pair of electrons occupying one dot with the states of the single electron occupying the other are used. The effective Hamiltonian model for single and two qubits was derived in Ref. \cite{Ferraro-2014} and in Ref. \cite{Ferraro-2015-qip}, respectively. For the single HY qubit the effective Hamiltonian is 
\begin{equation}\label{Hy}
H_{HY}=\frac{1}{2}E_z(\sigma_{1}^z+\sigma_{2}^z+\sigma_{3}^z)+\frac{1}{4}J\boldsymbol{\sigma}_{1}\cdot\boldsymbol{\sigma}_{2}+\frac{1}{4}J_{1}\boldsymbol{\sigma}_{1}\cdot\boldsymbol{\sigma}_{3}+\frac{1}{4}J_{2}\boldsymbol{\sigma}_{2}\cdot\boldsymbol{\sigma}_{3},
\end{equation} 
where $\boldsymbol{\sigma}_i$ ($i=1,2,3$) is the Pauli matrix of the i-th electron, $E_z$ is the Zeeman energy due to a constant global magnetic field used to initialize the qubit and the effective coupling constants $J_1$, $J_2$ and $J$ are explicitly derived in Ref. \cite{Ferraro-2014}. The key advantage of this qubit relies on the all-electrical manipulation of the qubit that assures very fast operations. 

The paper is organized as follows. Section 2 contains the presentation of the main results. The fidelities of the gate rotations $R_x(\theta)$ and $R_z(\theta)$ are calculated when the realistic transients of the control signal pulses are considered by adopting an appropriate filter function and the effect of the input disturbances is taken into account by using a Gaussian noise model. The rise and the fall edges of the realistic input signals are obtained by applying a first-order low-pass filter function to the ideal input signals. The low-pass filter with time constant $\tau=1/f_{max}$, where $f_{max}$ is the frequency cutoff, defines the bandwidth of the realistic input signal. Section 3 contains a discussion about the main findings obtained. Finally in Section 4 the nodal points related to the theory and the methods adopted are presented.
 
\section{Results}
We present for each spin qubit the results related to the single qubit gate operation $R_x(\theta)$ and $R_z(\theta)$ starting from the initial condition $|\psi(0)\rangle=\frac{1}{\sqrt{2}}(|0\rangle+i|1\rangle)$. The rotations are obtained through analytical input sequences that are reported in Appendix \ref{appendixA} \cite{Ferraro-2018}. We point out that the method is general and valid for arbitrary rotation angles as well as any initial condition. Moreover, the sequences are determined in such a way that each step time has to be longer than 100 ps, value that represents a current reasonable experimental limit.  

\subsection{$R_x(\theta)$ and $R_z(\theta)$ with bandwidth-limited pulses}
\begin{figure}[h]
\centering
\includegraphics[width=0.5\textwidth]{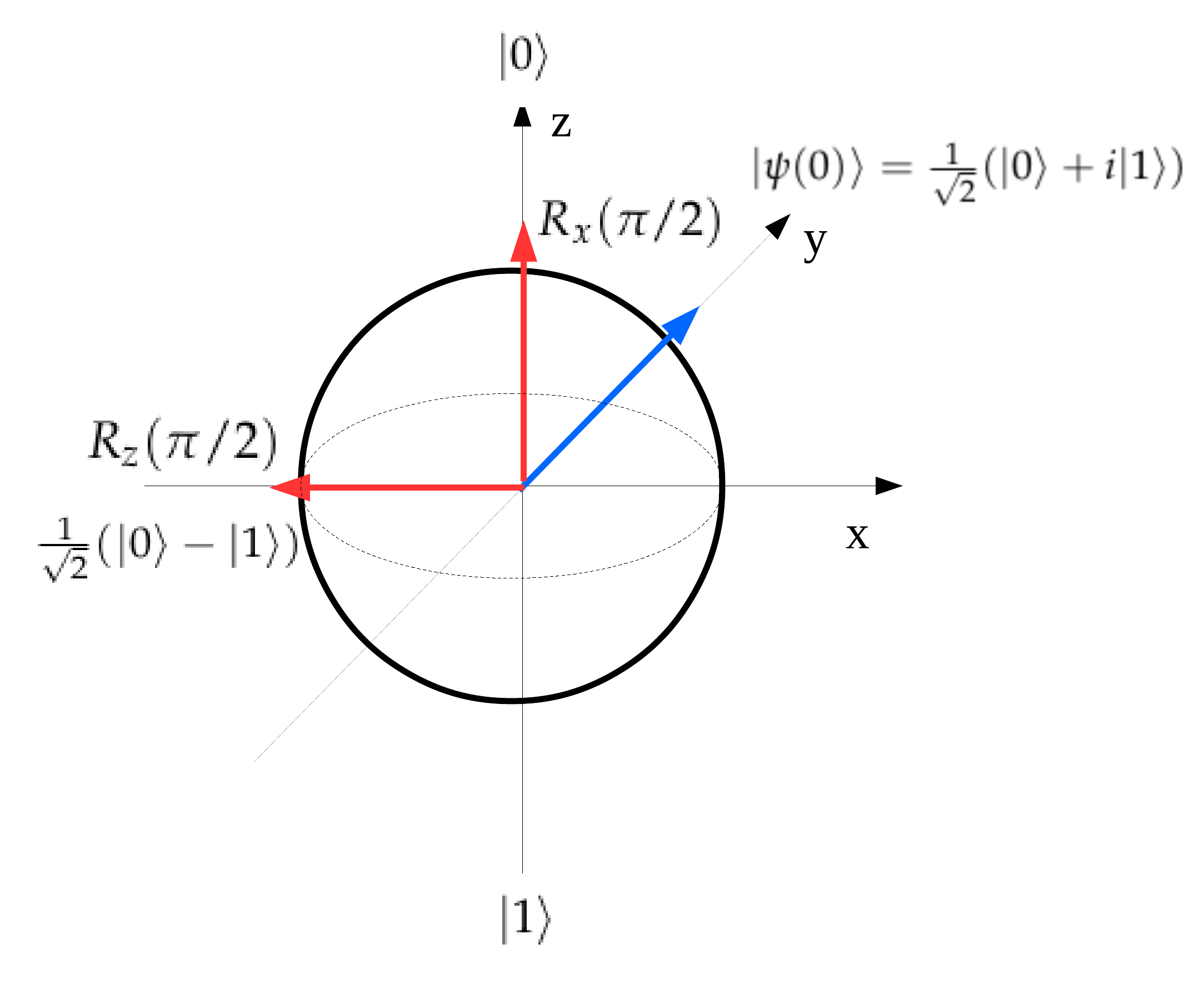}
\caption{Bloch sphere. The blue arrow represents the initial condition. The red arrows represent the final conditions after the $R_x(\pi/2)$ and $R_z(\pi/2)$ gate operations.}\label{Bloch}
\end{figure}

\unskip
Figure (\ref{Bloch}) shows a pictorial representation of the gate operations $R_x(\pi/2)$ and $R_z(\pi/2)$ on the Bloch sphere starting from the initial condition $|\psi(0)\rangle=\frac{1}{\sqrt{2}}(|0\rangle+i|1\rangle)$ represented by the blue arrow. The results of both the rotations are represented by the red arrows and the final state reached is explicitly written.

In the next subsections devoted to SS, ST and HY qubits respectively, we report gate infidelities as a function of the rotation angle $\theta$ and the time constant $\tau$ of first-order low-pass filter function without and with the effects of the input disturbances. 

\subsubsection{Single Spin qubit}
In the single spin qubit the signal sequences for the rotations along $x$ and $z$ axis differ from the number of the steps. While $R_x$ needs only one step, $R_z$ needs three steps where the $x$ and $y$ components of the oscillating magnetic field are obtained by modifying its phase $\phi$ (see Table \ref{table}). 

\begin{figure}[h]
\centering
\includegraphics[width=0.45\textwidth]{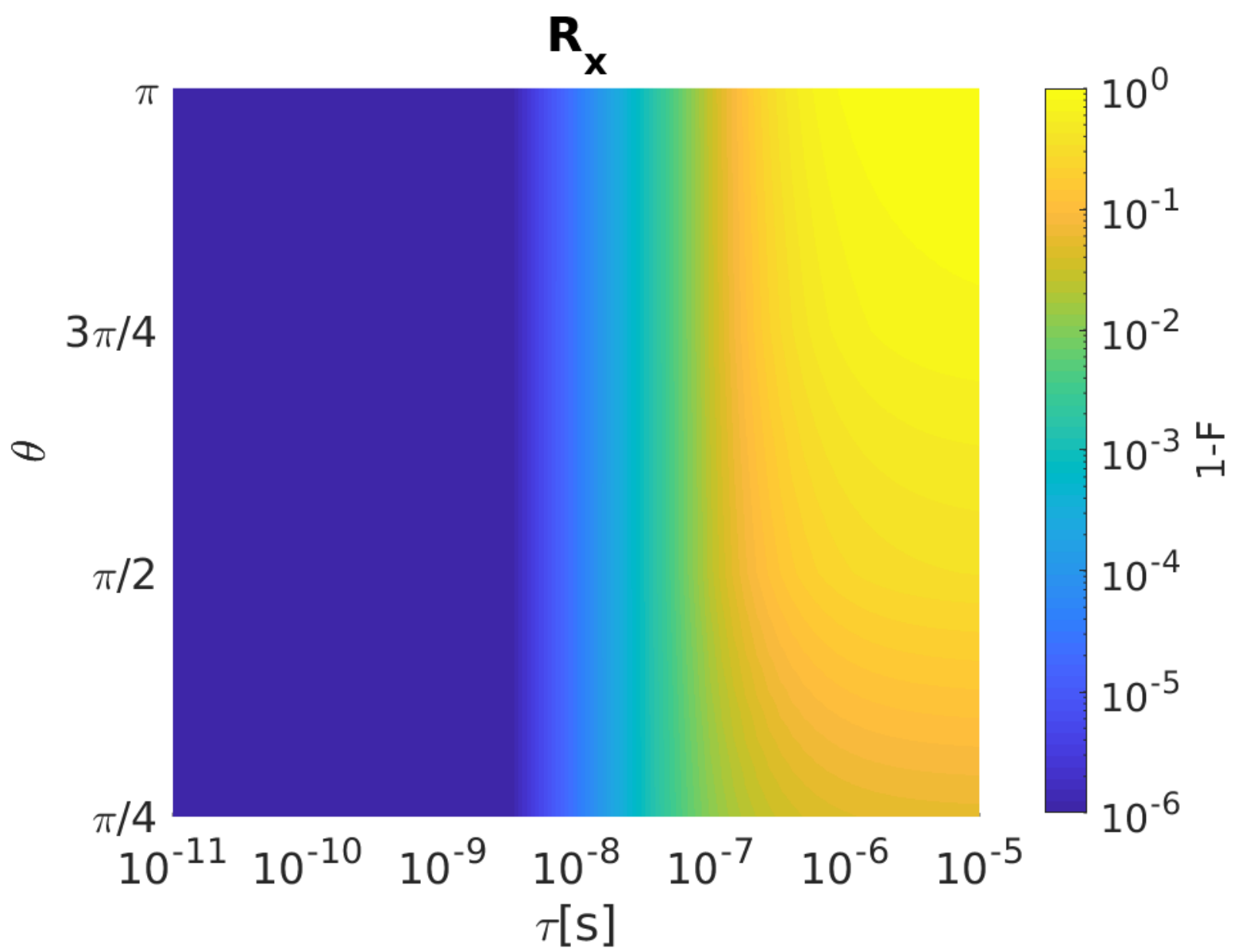}\ \includegraphics[width=0.45\textwidth]{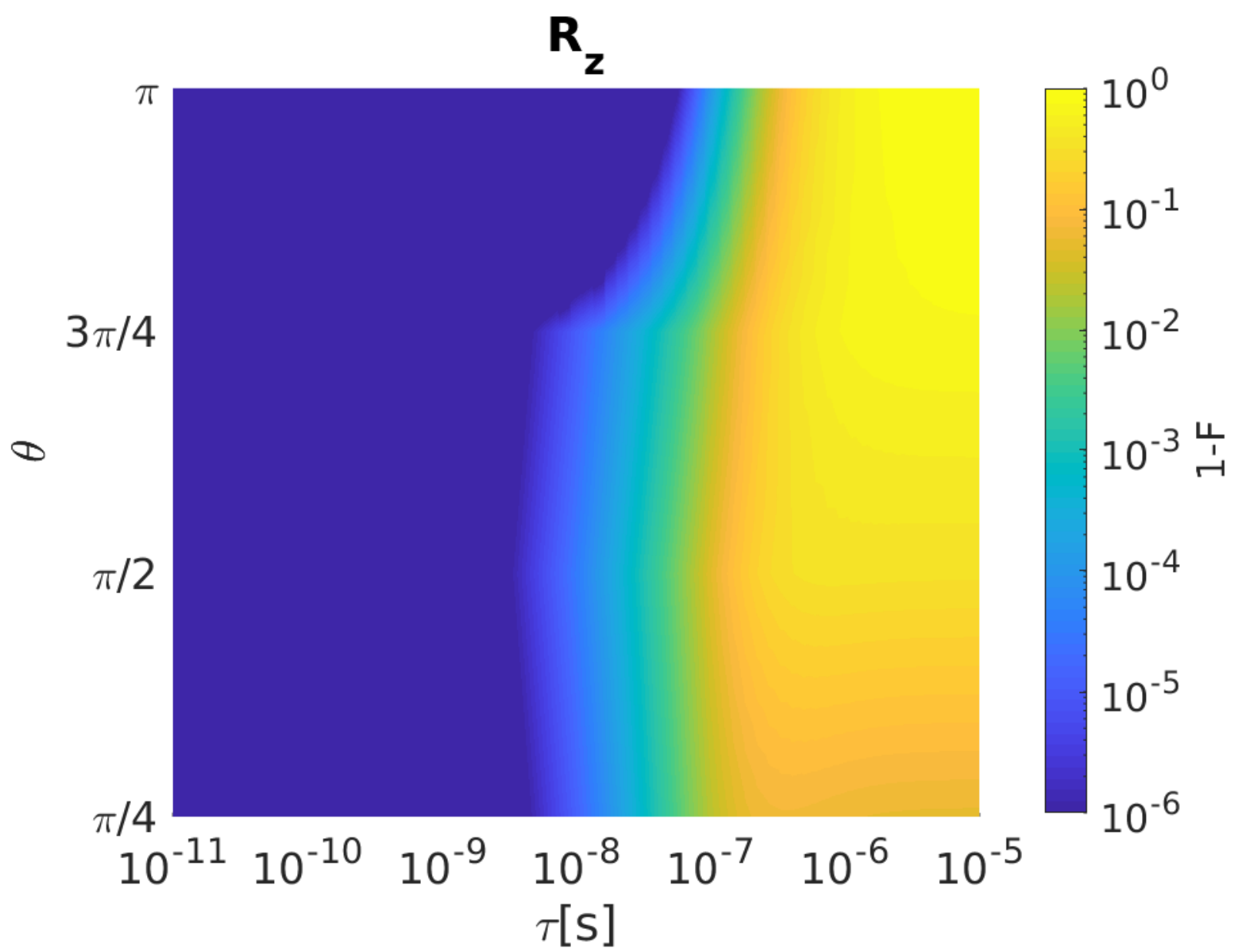}
\caption{SS qubit. Left: $R_x$ Infidelity as a function of $\theta$ and $\tau$ when bandwidth-limited input signals are considered. Right: The same for $R_z$. Qubit parameters: $\Omega/2\pi$=1 MHz, $\Delta\omega_z$= 0.}\label{SS1}
\end{figure}

In Figure (\ref{SS1}) we report $R_x$ (left) and $R_z$ (right) infidelity as a function of $\theta$ and $\tau$ when bandwidth-limited input signals are considered.
Both $R_x$ and $R_z$ infidelities increase as $\tau$ grows for all the considered rotation angles $\theta$. Note that $R_z$ fidelity degrades slowly for $\theta=\pi$ than other rotation angles as $\tau$ increases.

\begin{figure}[h]
\centering
\includegraphics[width=0.45\textwidth]{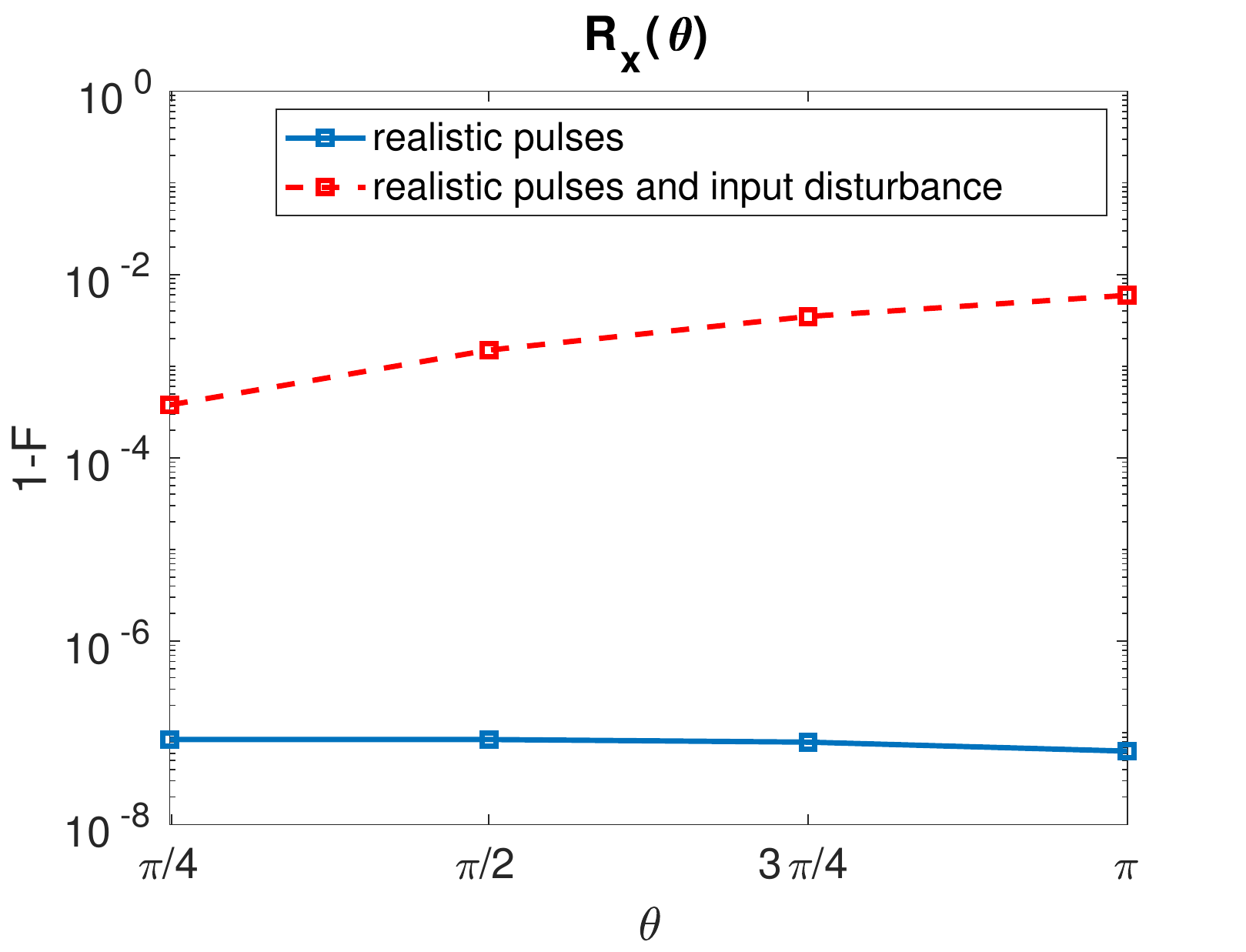}\ \includegraphics[width=0.45\textwidth]{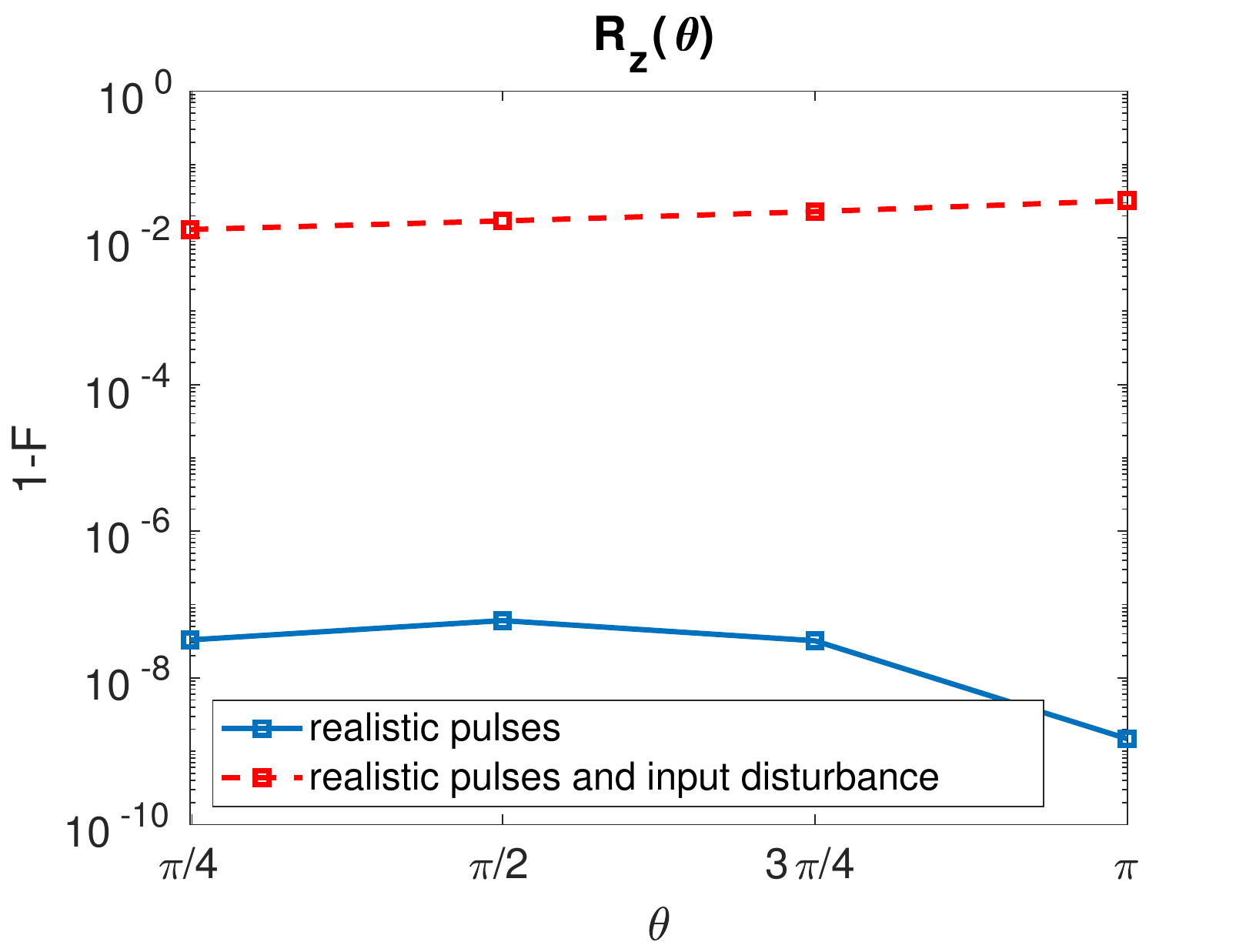}
\caption{SS qubit. Left: $R_x$ infidelity as a function of $\theta$ when undisturbed input signals (solid line, blue) and disturbed input signals with $\sigma_{\Omega/2\pi}$= 0.05 MHz \cite{Kawakami-2014} and $\sigma_{\Delta\omega_z/2\pi}$= 20 Hz \cite{DatasheetMWgenerator} (dashed line, red) are considered. The value of $\tau$ is fixed to 100 ps. Right: The same for $R_z$.}\label{SS2}
\end{figure}

Inclusion of the input signal disturbances in our calculations gives the results reported in Figure (\ref{SS2}) for the SS qubit. Here, by using a filter time constant of $\tau=100$ ps we present $R_x$ (left) and $R_z$ (right) infidelity as a function of $\theta$ when undisturbed (solid line, blue) and disturbed (dashed line, red) input signals are considered. 
Both $R_x$ and $R_z$ fidelities are heavily deteriorated by the input disturbances. Disturbed $R_x$ shows an increased infidelity as $\theta$ grows due to the fact that control sequence times for large $\theta$ are longer than those for small $\theta$. As a result, input disturbances are integrated for longer times and gate fidelity worsen. The same comment holds for $R_z$ but with a mild infidelity increase as $\theta$ augments.     

\subsubsection{Singlet-Triplet qubit}
In the singlet-triplet qubit the signal sequences for the rotations along $x$ and $z$ axis are reported in Table \ref{table}. $R_x$ is obtained operating in one step with the input $\Delta E_z$, while $R_z$ needs two steps that include also the manipulation of the exchange coupling $J$. 

\begin{figure}[h]
\centering
\includegraphics[width=0.45\textwidth]{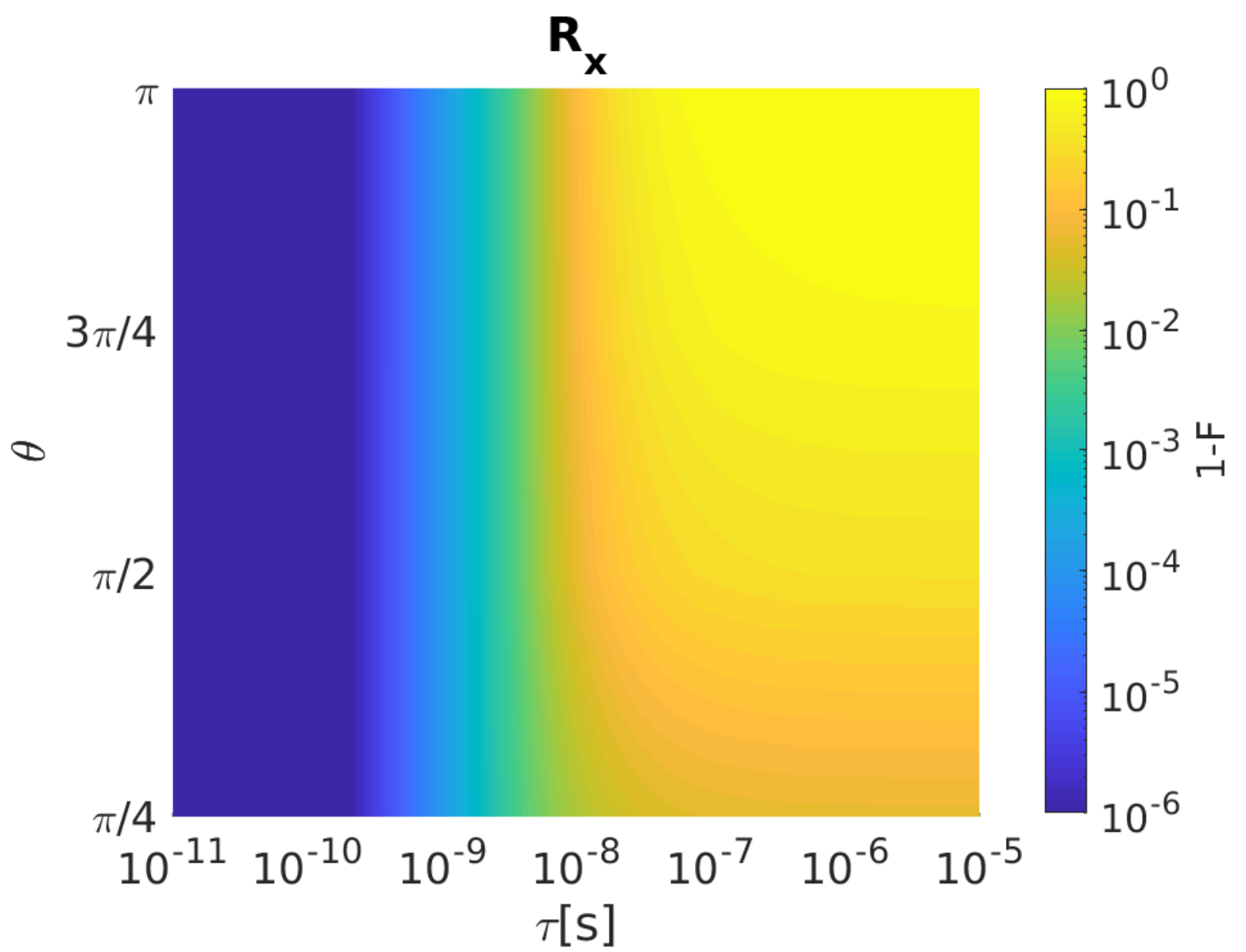}\ \includegraphics[width=0.45\textwidth]{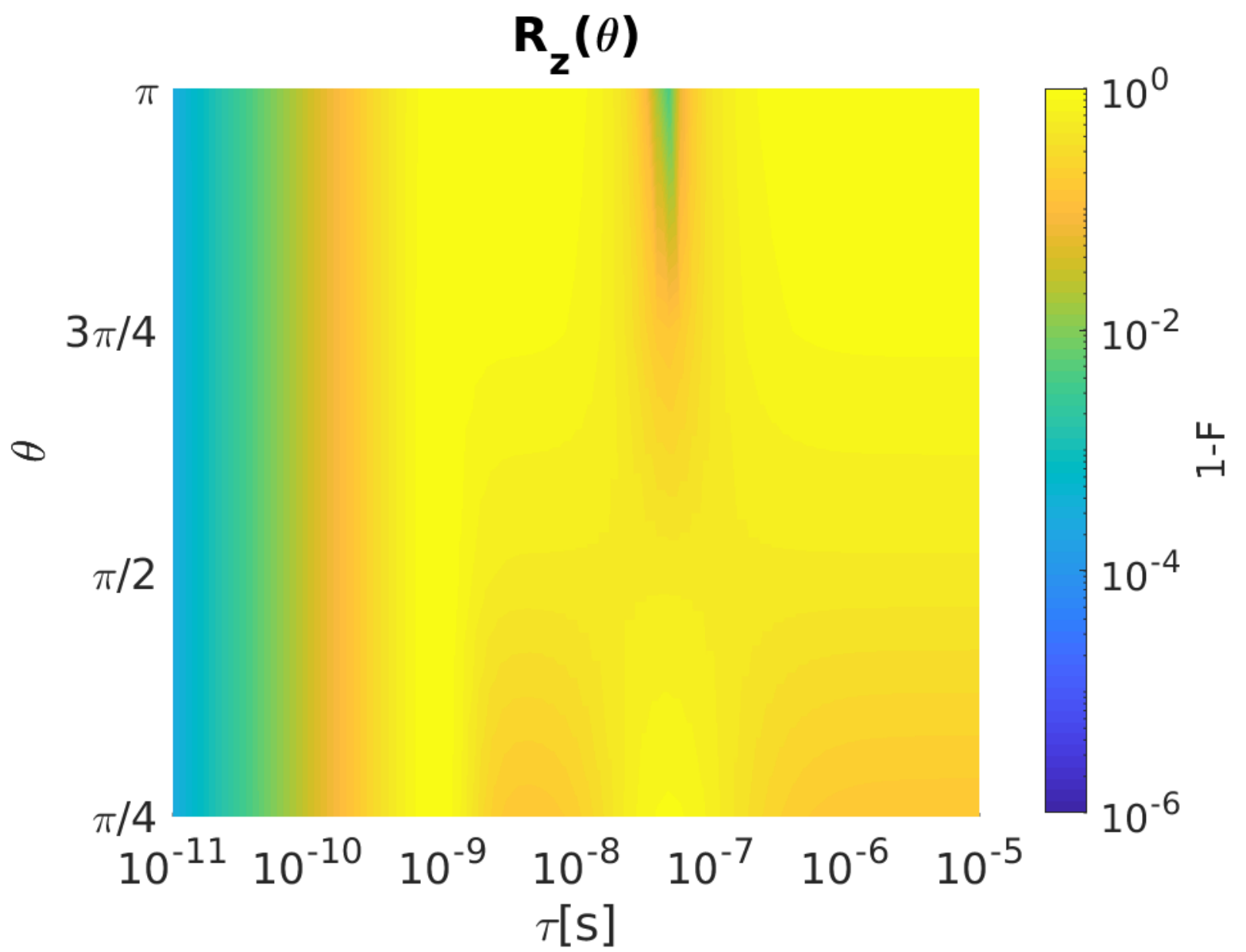}
\caption{ST qubit. Left: $R_x$ Infidelity as a function of $\theta$ and $\tau$ when bandwidth-limited input signals are considered. Right: The same for $R_z$. Qubit parameters: $J$=700 neV, $\Delta E_z$= 32 neV \cite{XianWu-2014}.}\label{ST1}
\end{figure}

\begin{figure}[h]
\centering
\includegraphics[width=0.45\textwidth]{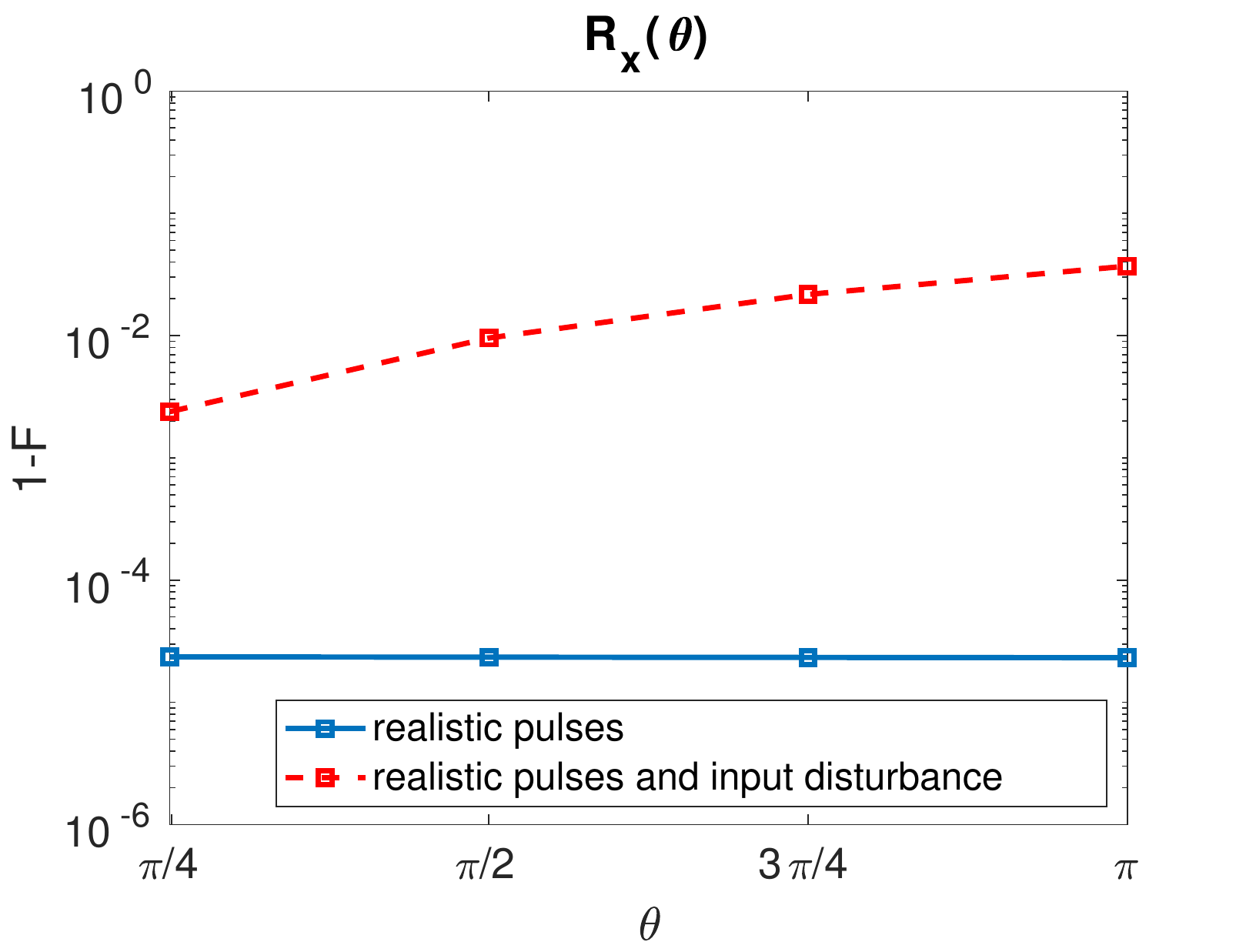}\ \includegraphics[width=0.45\textwidth]{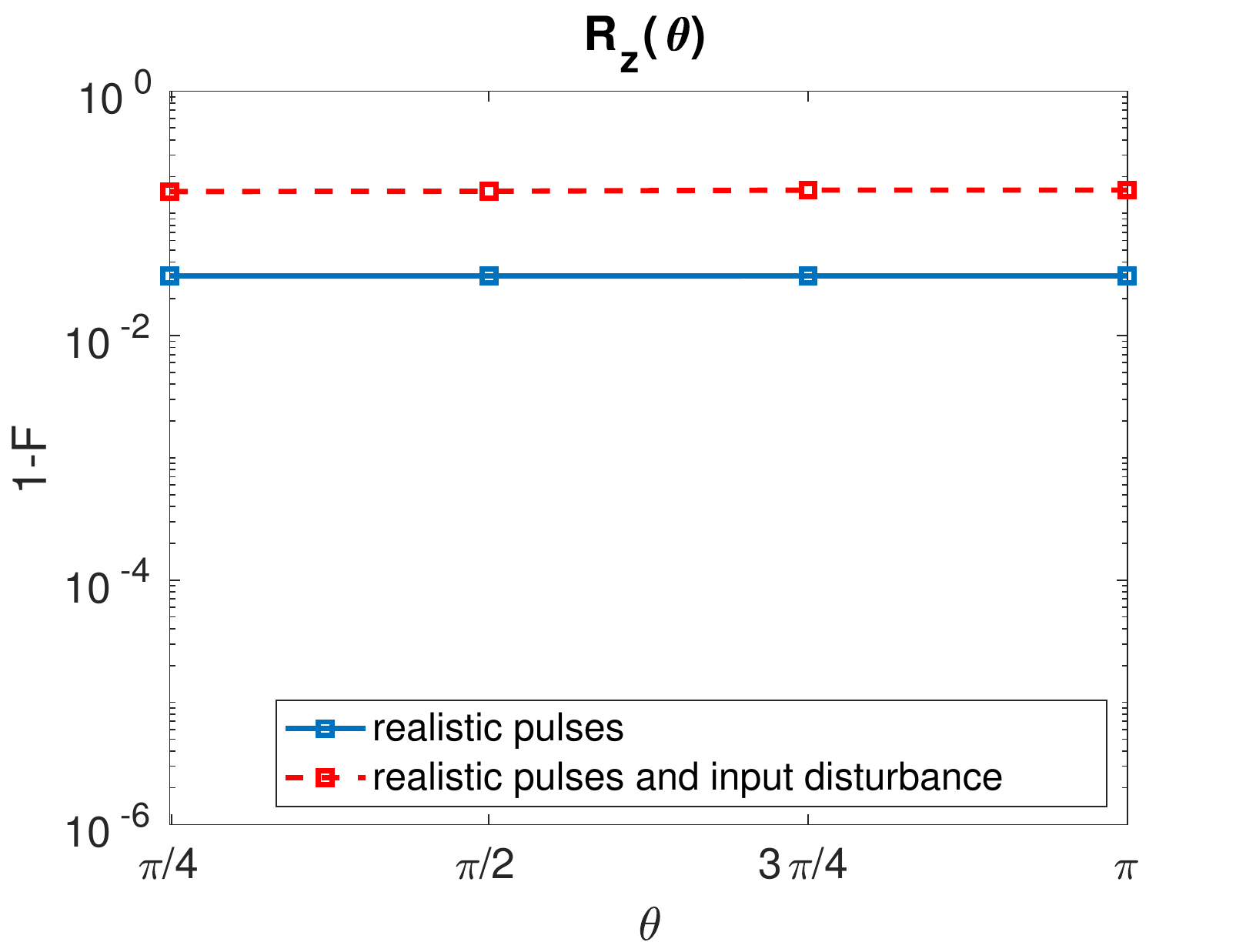}
\caption{ST qubit. Left: $R_x$ Infidelity as a function of $\theta$ when undisturbed input signals (solid line, blue) and disturbed input signals with $\sigma_J$= 1 neV, $\sigma_{\Delta E_z}$= 4 neV \cite{XianWu-2014} (dashed line, red) are considered. The value of $\tau$ is fixed to 100 ps. Right: The same for $R_z$.}\label{ST2}
\end{figure}

Figure (\ref{ST1}) shows $R_x$ (left) and $R_z$ (right) infidelity as a function of $\theta$ and $\tau$.
$R_x$ and $R_z$ infidelities increase as $\tau$ grows for all the considered rotation angles. After the inclusion of the input disturbances, the resulting ST qubit infidelities for both rotations with $\tau=100$ ps are reported in Figure (\ref{ST2}). 
As for the single spin qubit, singlet-triplet qubit rotations $R_x$ and $R_z$ show a strong fidelity degradation when input disturbances are included. $R_x$ infidelity grows as $\theta$ increases whereas $R_z$ infidelity is not sensitive to $\theta$ variations. 

\subsubsection{Hybrid qubit}
The rotations along $x$ and $z$ axis for the hybrid qubit are realized through signal sequences involving the effective exchange couplings $J$, $J_1$ and $J_2$. They are multi-step sequences composed respectively by two and three steps (see Table \ref{table}). 

\begin{figure}[h]
\centering
\includegraphics[width=0.45\textwidth]{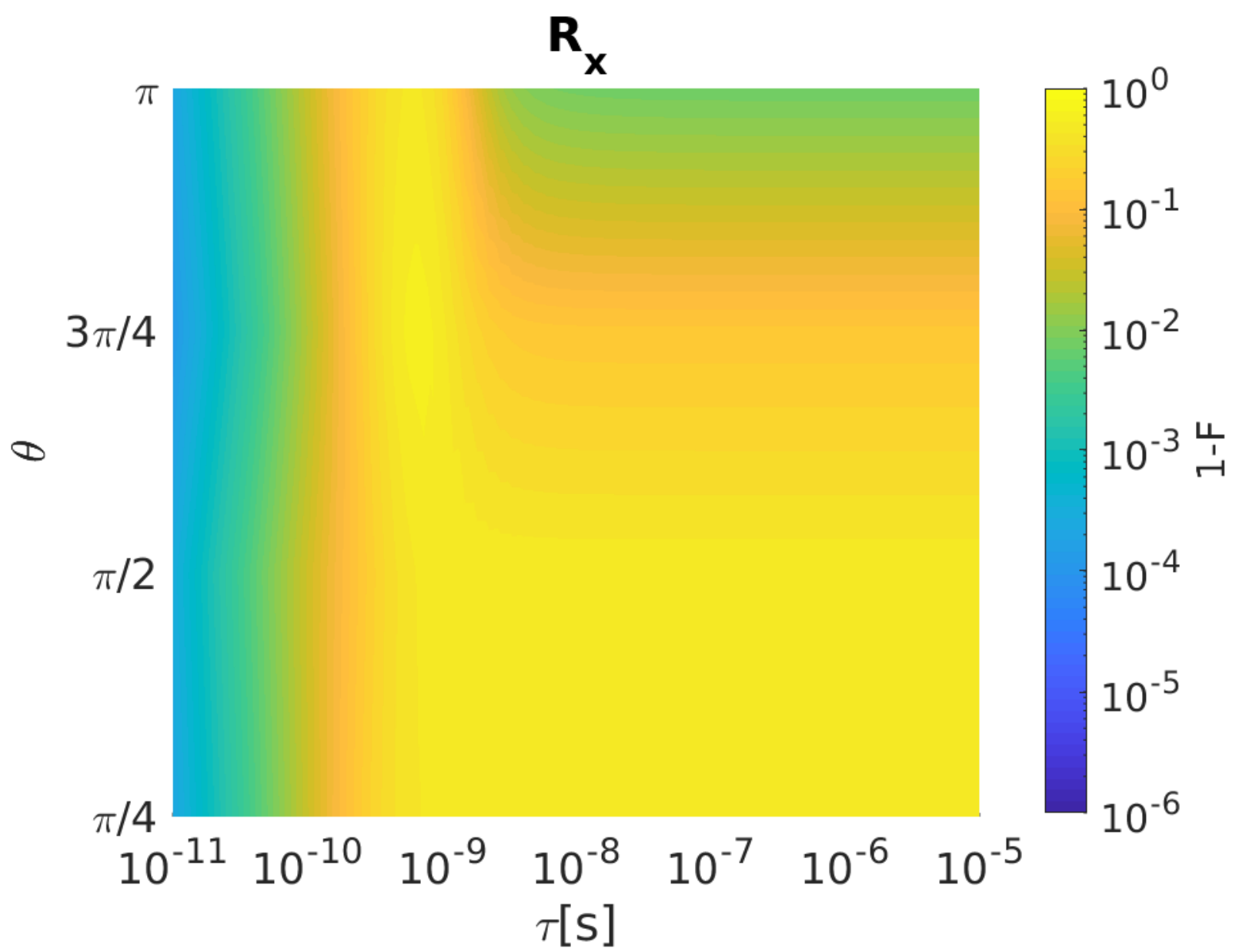}\ \includegraphics[width=0.45\textwidth]{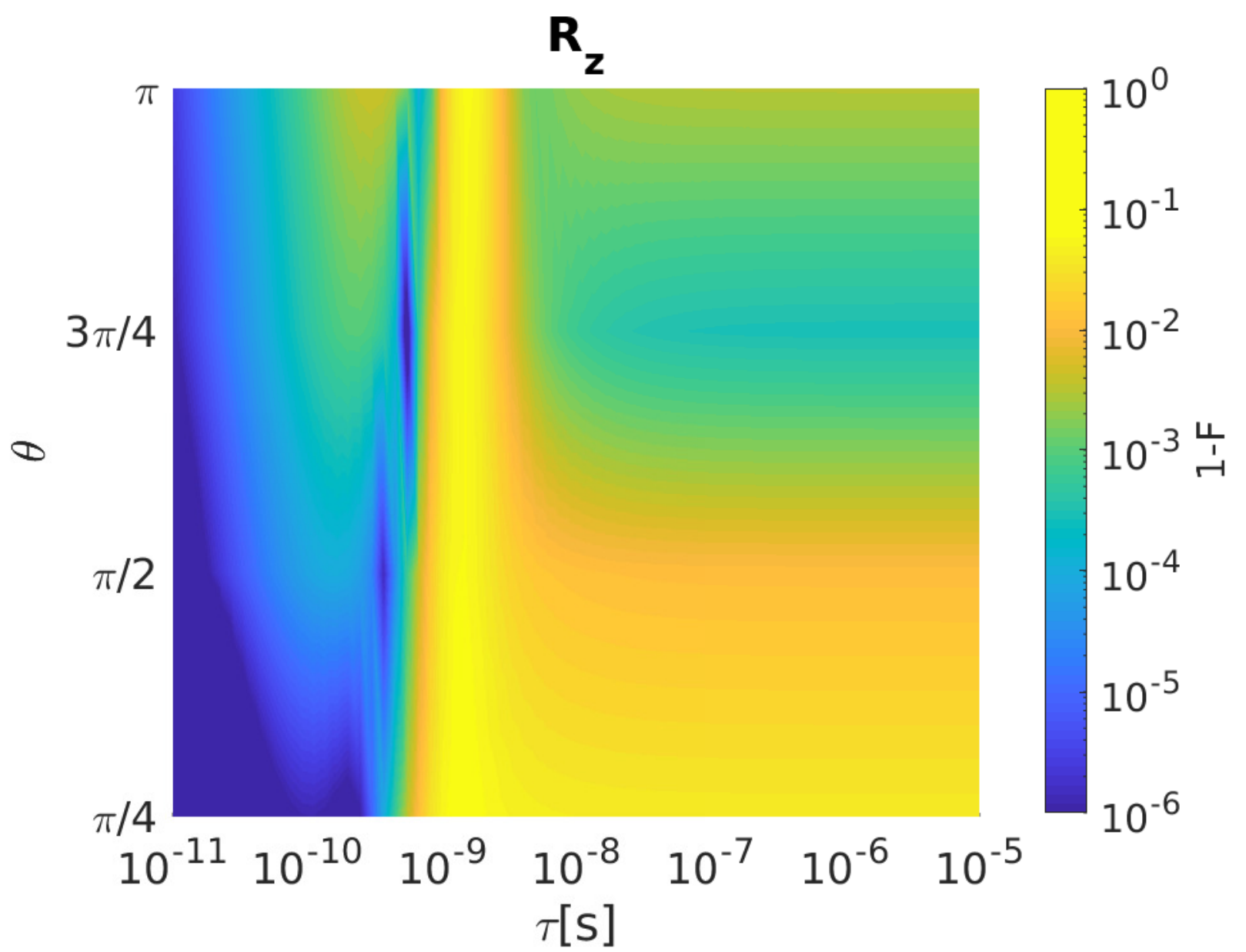}
\caption{HY qubit. Left: $R_x$ Infidelity as a function of $\theta$ and $\tau$ when bandwidth-limited input signals are considered. Right: The same for $R_z$. Qubit parameters: $J_1$=$J_2$=1 $\mu$eV, $J$= 0.5 $\mu$eV \cite{DeMichielis-2015}.}\label{HY1}
\end{figure} 
As it is evident in Figure (\ref{HY1}), $R_x$ infidelity increases as $\tau$ grows for almost all the considered rotation angles. An infidelity reduction is observable for $\theta$=$\pi$ in correspondance to $\tau$>1 ns. $R_z$ infidelity instead shows a more complex behaviour than $R_x$. 1-F increases for all $\theta$ as $\tau$ augments till roughly 100 ps. When $\tau$ is set between 100 ps and 1 ns some local minima in the infidelity can be observed at different $\theta$.  
For $\tau$> 1 ns, infidelity is very high (larger than $10^{-2}$) and constant below $\theta$=$\pi$/2 whereas it decreases for rotation angles above $\pi$/2.

\begin{figure}[h]
\centering
\includegraphics[width=0.45\textwidth]{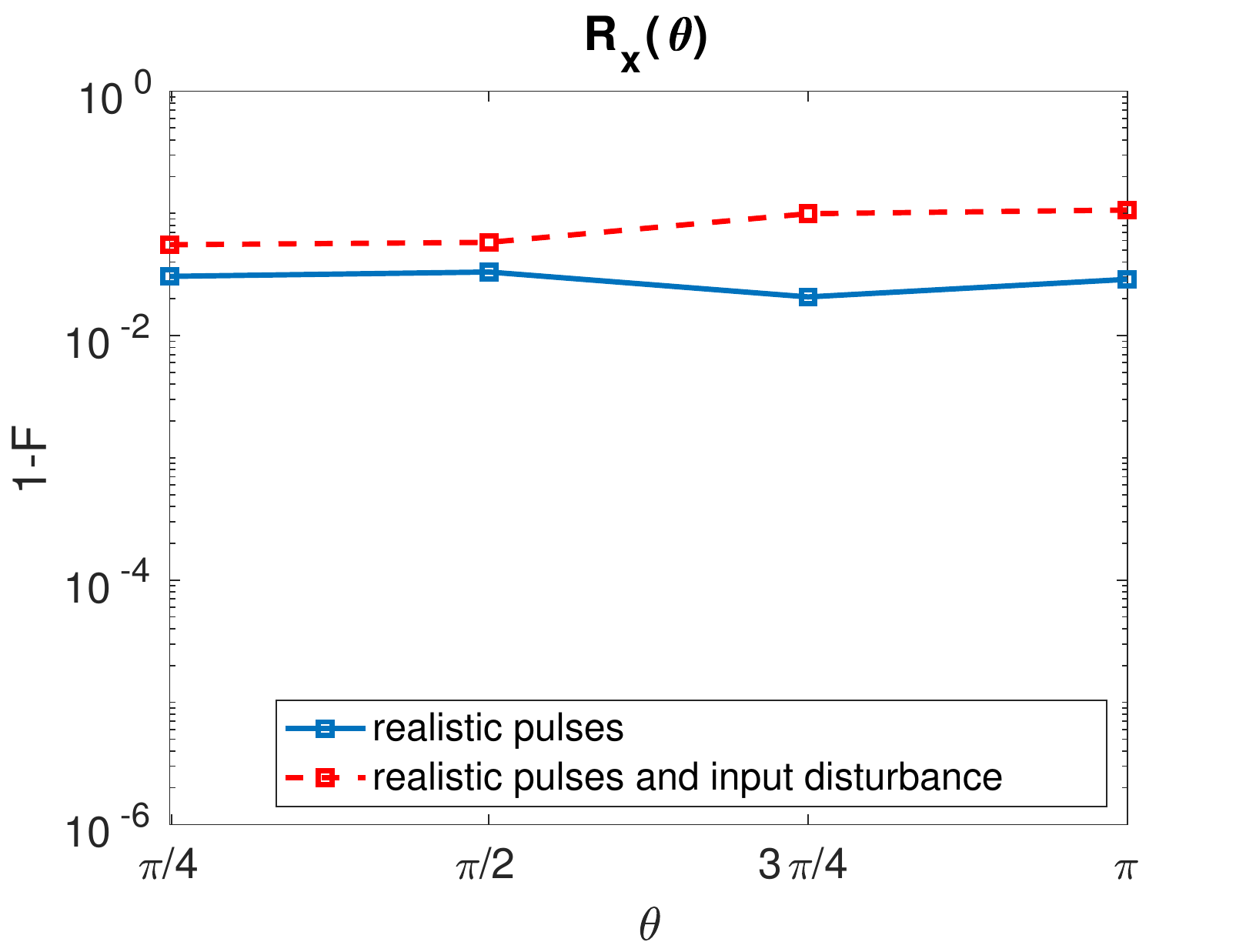}\ \includegraphics[width=0.45\textwidth]{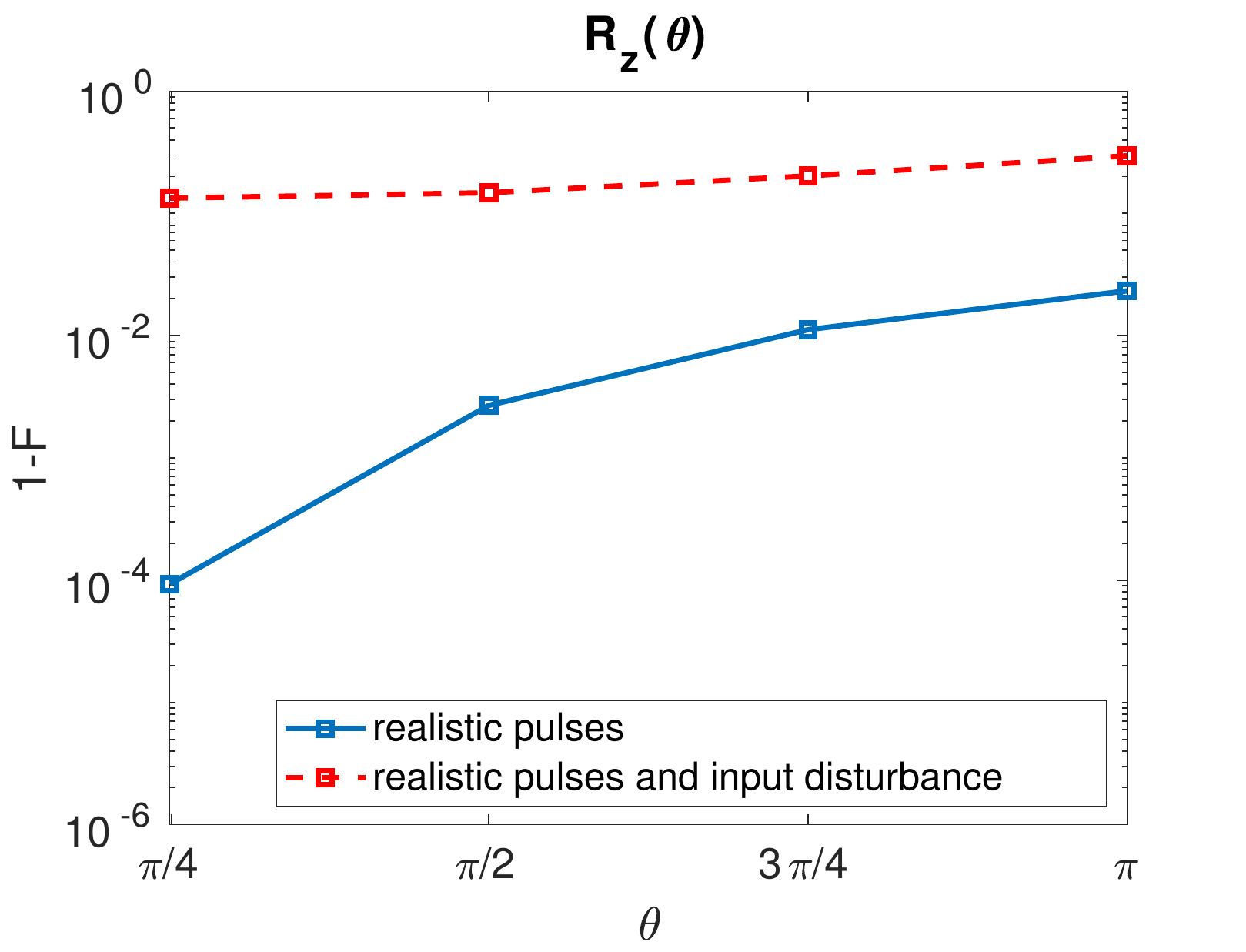}
\caption{HY qubit. Left: $R_x$ Infidelity as a function of $\theta$ when undisturbed input signals (solid line, blue) and disturbed input signals with $\sigma_{J}$= 80 neV \cite{Thorgrimsson-2017} (dashed line, red) are considered. The value of $\tau$ is fixed to 100 ps. Right: The same for $R_z$.}\label{HY2}
\end{figure}

After inclusion of the input disturbances, the resulting HY qubit infidelities for both rotation with $\tau=100$ ps are reported in Figure (\ref{HY2}). 
Hybrid qubit rotation $R_x$ shows a weak fidelity degradation with respect to the undisturbed case, in addition a slight degradation of the fidelity can be observed as $\theta$ increases. Conversely, $R_z$ infidelity is strongly affected by the input disturbances in the entire range of $\theta$ studied with a weak additional degradation for large $\theta$.      

\subsection{Fidelity comparison}
\begin{figure}[h]
\centering
\includegraphics[width=0.45\textwidth]{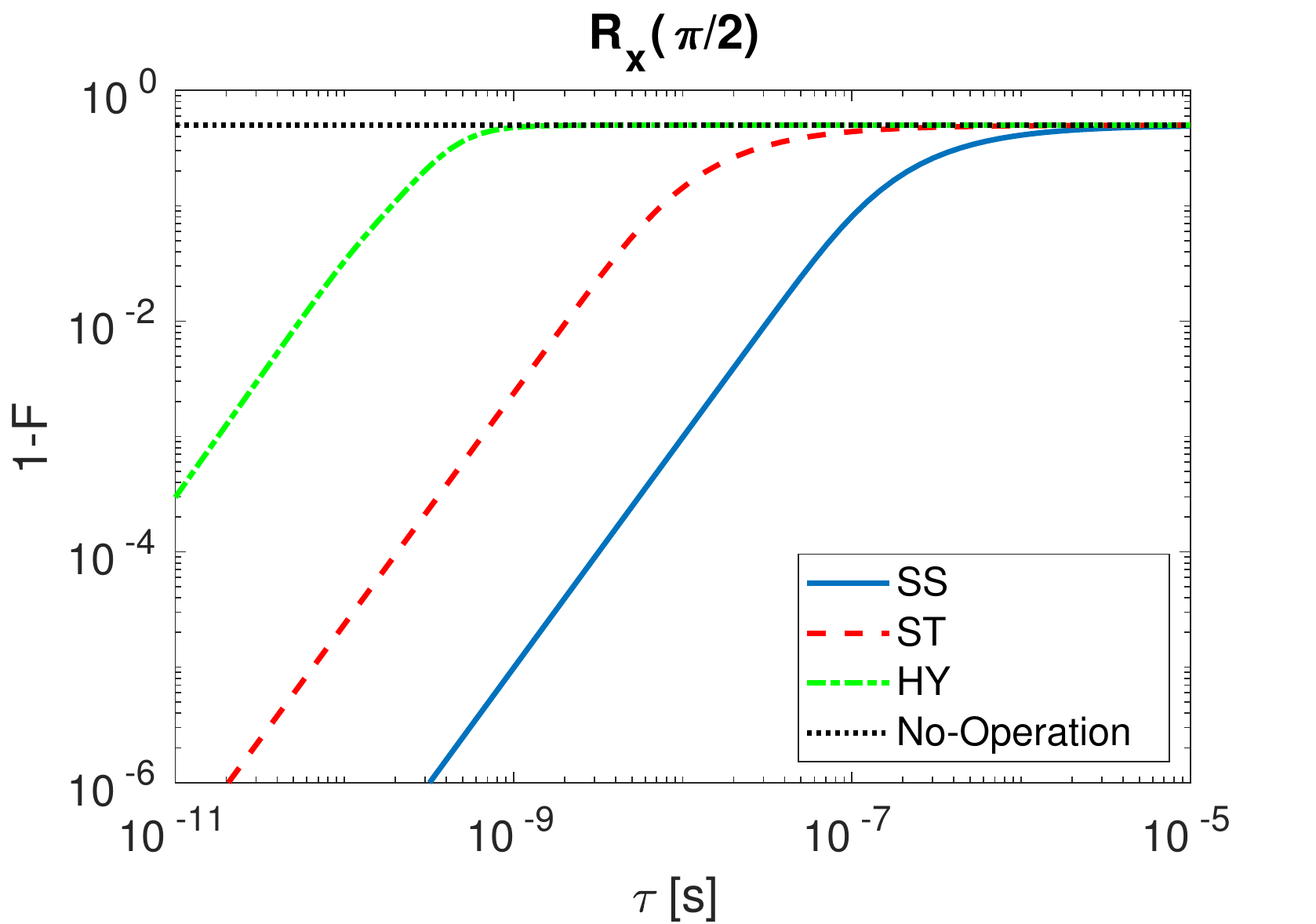}\ \includegraphics[width=0.45\textwidth]{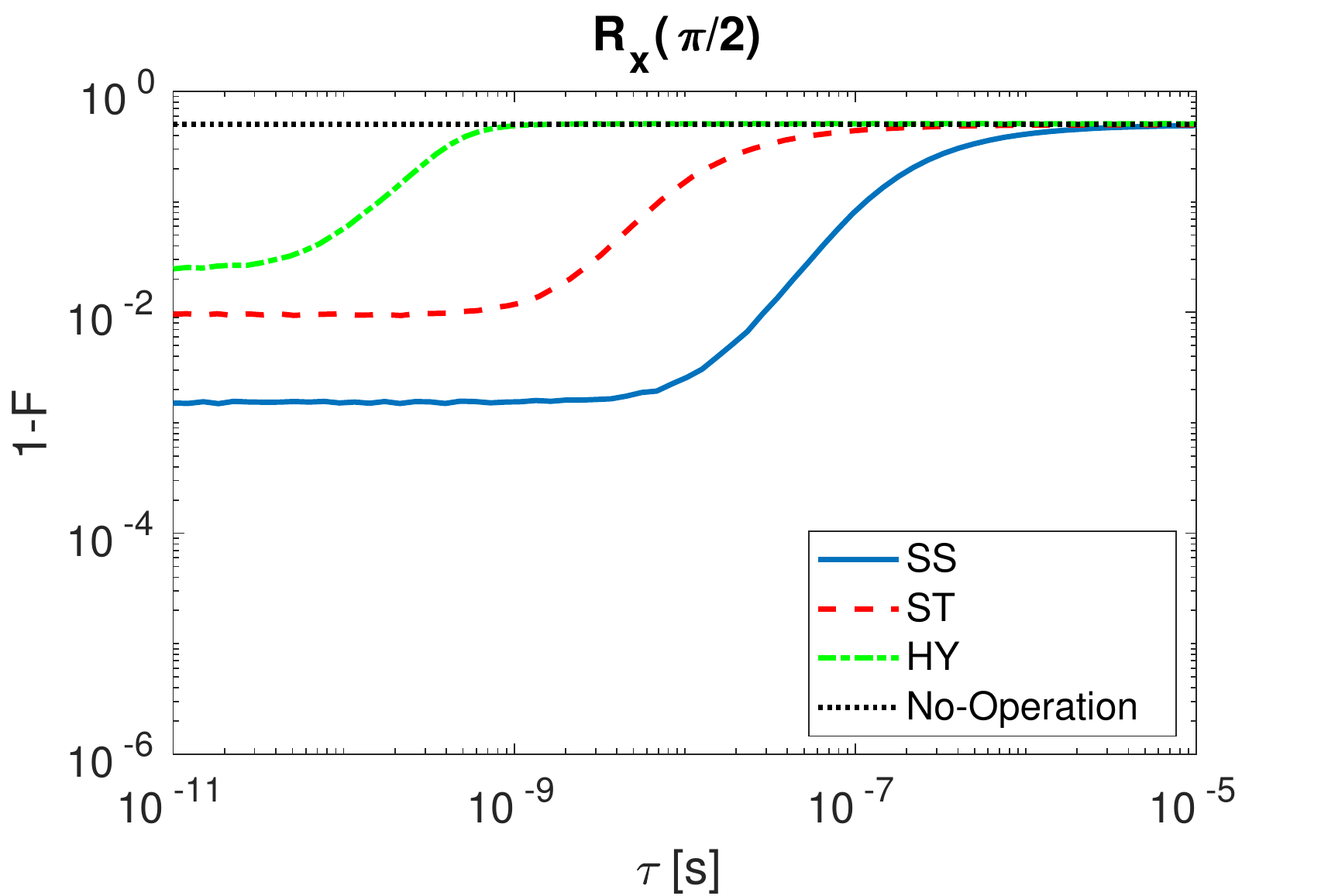}
\caption{$R_x(\pi/2)$ infidelity as a function of $\tau$ for SS (solid line, blue), ST (dashed line, red), HY (dot-dashed line, green) qubit and No-Operation (dotted line, black). Left: with undisturbed input signals. Right: with disturbed input signals.}\label{F1}
\end{figure}

\begin{figure}[h]
\centering
\includegraphics[width=0.45\textwidth]{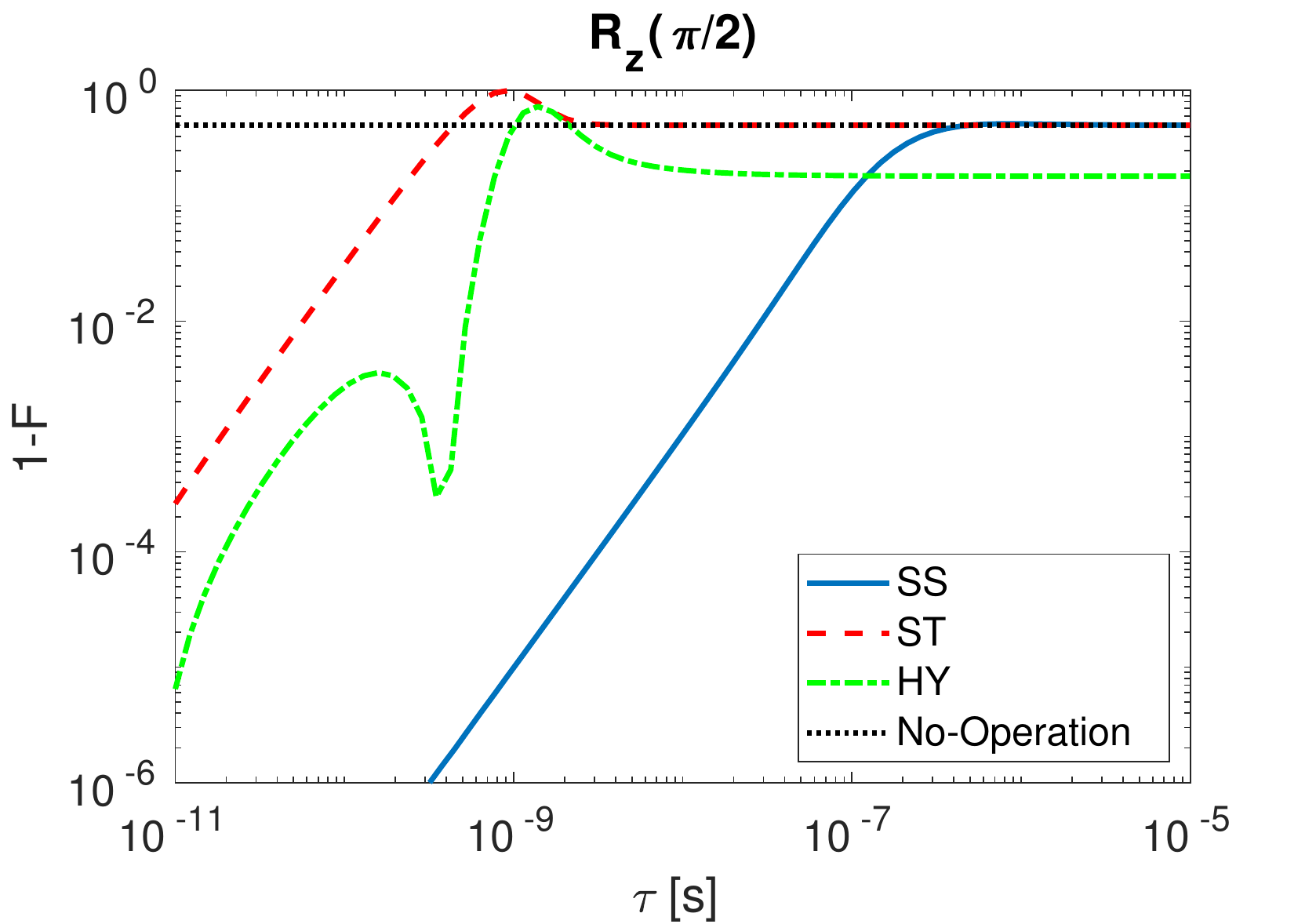}\ \includegraphics[width=0.45\textwidth]{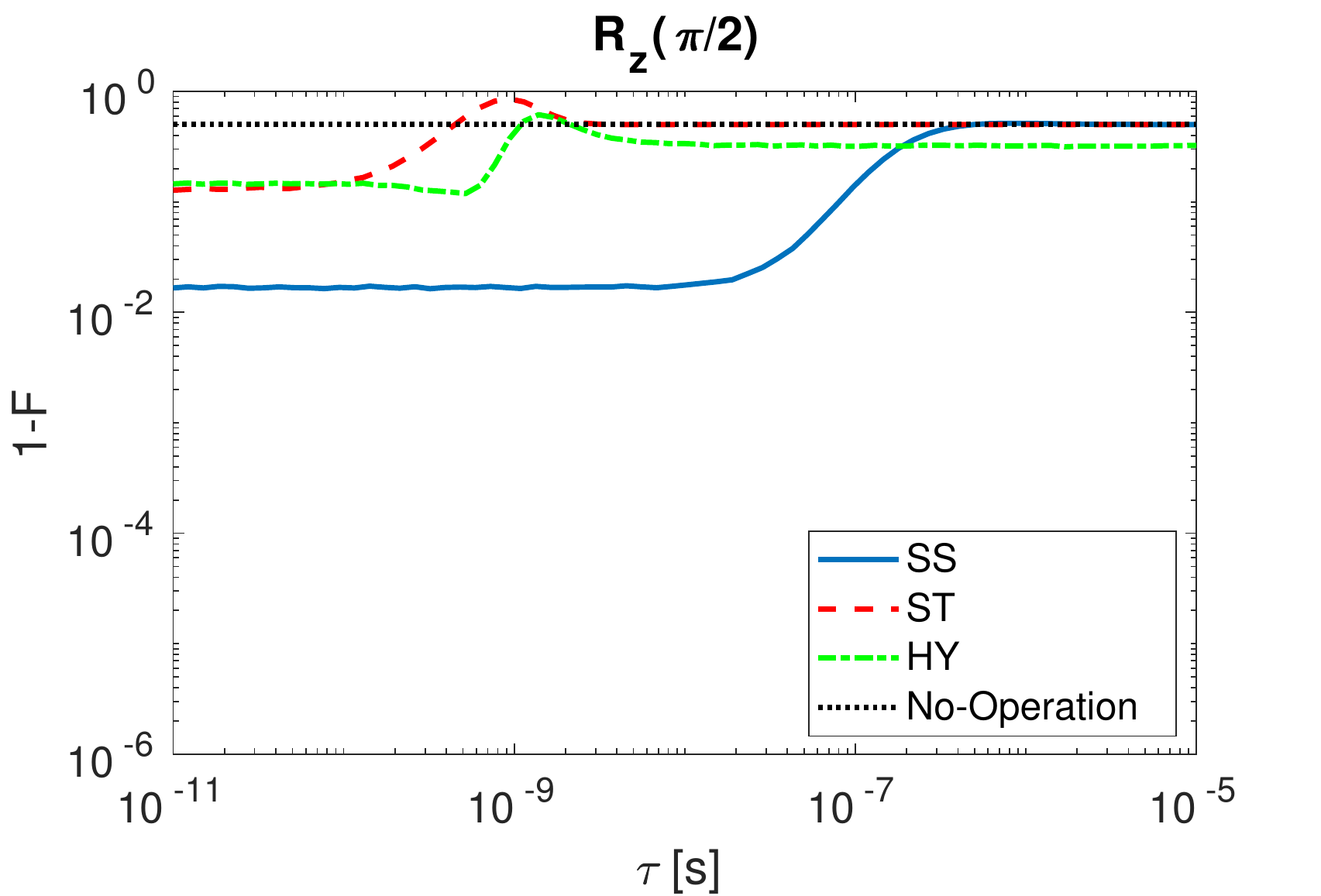}
\caption{$R_z(\pi/2)$ infidelity as a function of $\tau$ for SS (solid line, blue), ST (dashed line, red), HY (dot-dashed line, green) qubit and No-Operation (dotted line, black). Left: with undisturbed input signals. Right: with disturbed input signals.}\label{F2}
\end{figure}

A comparison of the three spin qubit under study is here shown. The infidelity behaviour as a function of $\tau$ is analized for $R_x(\pi/2)$ (Fig. (\ref{F1})) and $R_z(\pi/2)$ (Fig. (\ref{F2})). The results are presented when undisturbed pulses are considered (left) and when input disturbances are included (right). For both the operations we observe that the fidelity decreases when $\tau$ grows.

From Fig. (\ref{F1}), we may conclude that, setting the same initial condition for the gate operation considered, the SS qubit assures greater fidelities with respect to ST and HY qubit. We observe for the HY which has shorter sequences that is instead the most sensitive to $\tau$ variation. In the case in which also the input disturbance is included (right) we observe for SS and ST a saturated behavior when $\tau$ is reduced, also the HY fidelity saturates but for smaller values of $\tau$. The No-Operation curve describes the case in which the realistic pulses to perform the rotations are not applied. In this case the fidelity gives as a result $1/2$, due to the reciprocal position of the initial and the ideal final qubit states on the Bloch sphere. When tau is large the time variation of the control signal is so tiny that the qubit state is minimally rotated from the initial condition and thus we observe a saturated behavior of the infidelity to the value corresponding to No-Operation as tau increases.  

The situation changes for $R_z(\pi/2)$ (Fig. (\ref{F2})) that is obtained with multiple step sequences (see Table \ref{table}). SS qubit is once again the most robust to $\tau$ variation, but differently from $R_x(\pi/2)$, the most sensitive is the ST qubit. The HY qubit present local minima in the gate infidelity. Those local minima originate because selected gate sequences are not the shortest ones in absolute but they have to last longer than 100 ps. Such a step time elongation is obtained by increasing the parameter $n$ (see Table \ref{table}) causing the qubit state to make additional complete rotations on the Bloch sphere during the step. When input signal sequences of such kind are filtered at a given $\tau$, the consequential delay in the signal switching (on and off) generates partial rotations on the Bloch sphere that if sum up to a 2$\pi$ rotation can lead to operations with low infidelity. The presence of an infidelity maximum in the ST qubit before the saturation to the No-Operation infidelity value (dotted line, black) is strictly connected to the nature of the multi-steps operation. The HY presents a similar behavior except from the fact that the infidelity does not reach the No-Operation value in the range of $\tau$ studied, since the $R_z(\pi/2)$ operation requires that the exchange coupling $J$, that is not tunable from external gates, is always turned on during all the operations.

\section{Discussion}
As expected, when bandwidth-limited input signals for $x$ and $z$ rotations are considered, all the qubit types show a reduction of gate fidelities for increasing $\tau$. Inclusion of signal amplitude disturbances further deteriorates the gate fidelity for reduced $\tau$, also creating plateaus and leading to fidelities values in the range between 90\% and 99.99\%. 
The presence of local minima in the gate infidelities of HY qubit suggests that optimal working points can be identified achieving not only a reduced infidelity but even a simultaneous relaxation of the bandwidth requirements of the input system (larger $\tau$). By using parameter values at the state of the art we can conclude that the hybrid qubit has the lowest infidelity provided that input signals with large enough bandwidth are available to achieved those fast sequences. Conversely, the single spin qubit shows low infidelities even at relatively quite high time constants due to the slow pulse times of its sequences.

\section{Methods}

The qubit dynamics is obtained solving the master equation $\frac{\partial\rho}{\partial t}=-\frac{i}{\hbar}[H,\rho]$ for the total density matrix $\rho$, where $H$ is the effective qubit Hamiltonian ($H\equiv H_{SS},H_{ST},H_{HY}$) in the logical basis $\{|0\rangle,|1\rangle\}$. The ideal gate sequences for $R_x(\theta)$ and $R_z(\theta)$ are analytically derived and reported in Appendix \ref{appendixA}.

The solution of the ideal dynamics, in which the applied pulses have ideal rise and fall edges (squared signals), is compared with the realistic situation in which the rise and the fall edges of the input signals are described by a first-order low-pass filter function with time constant $\tau$.

Moreover also the non-idealities of the input amplitudes are included. Employing the quasi-static model, the errors on the input signals are modeled as random variables with Gaussian distributions featuring zero mean and standard deviation $\sigma$ that add up to the ideal values. The figure of merit used to estimate the disturbance effects is the fidelity $F=\left[\Tr\sqrt{\sqrt{\rho^{ideal}}\rho^{real}\sqrt{\rho^{ideal}}}\right]^2$ that measure how much the real state is distant from the ideal case.

\section*{Acknowledgments}
This work has been funded from the European Union's Horizon 2020 research and innovation programme under grant agreement No 688539.

\appendix
\section{Analytical gate sequences}\label{appendixA}
\unskip
Table \ref{table} contains the analytical gate sequences for all the spin qubit under study that realize in one-step or multiple-steps the rotations $R_x(\theta)$ and $R_z(\theta)$ on the Bloch sphere \cite{Ferraro-2018}.
\begin{table}[h]
\caption{Analytical gate sequences that realize $R_x(\theta)$ (second column) and $R_z(\theta)$ (third column). Each row refers to a qubit type.}\label{table}
\centering
\begin{tabular}{cp{5cm}p{5cm}}
\hline
Qubit & \textbf{$R_x(\theta)$}	& \textbf{$R_z(\theta)$}	\\
\hline
SS & $t_x(\theta)=\frac{\theta}{\Omega_x}$
& $t_y=(-\pi/2)/\Omega_y\hspace{2cm}$ $t_x(\theta)=\theta/\Omega_x\hspace{2.5cm}$ $t_y=(\pi/2)/\Omega_y$ \\
\hline
ST & $t_z(\theta)=\left(\frac{\theta}{4\pi}+n\right)\frac{h}{\Delta E_z}\hspace{1cm}$ with $n$ integer
& $t_z=\frac{n}{2}\frac{h}{\Delta E_z}$ $\hspace{3cm}$ $t_{J}(\theta)=\left(-\frac{\theta}{2\pi}+n\right)\frac{h}{J}$\\
\hline
HY & $t_{J_1}(\theta)=\left(\frac{n}{C}-\frac{1}{\sqrt{3}}\frac{\theta}{2\pi}\frac{1}{J^{max}}\right)h\hspace{0.1cm}$  $t_{J_2}(\theta)=\left(\frac{n}{C}+\frac{1}{\sqrt{3}}\frac{\theta}{2\pi}\frac{1}{J^{max}}\right)h\hspace{0.1cm}$ with $n=\left\lceil \frac{C}{J^{max}}\frac{1}{\sqrt{3}}\frac{\theta}{2\pi} \right\rceil,\hspace{1cm}$ $C=E_z+\frac{3}{4}J^{max}$ and $\hspace{2cm}J^{max}=\max{(J_1)}=\max{(J_2)}$ 
& $t_{J_1}(\theta)=\frac{1}{C}\left[\frac{\theta}{\pi}A+sign\left(\frac{2\pi}{3}-\theta\right)B\right]\frac{h}{J^{max}}\hspace{0cm}$  $t_{J_2}(\theta)=t_{J_1}(\theta)\hspace{2cm}$ $t_{J}(\theta)=\left(2-\frac{\theta}{\pi}\right)\frac{h}{J^{max}}\hspace{1cm}$ with $A=\frac{E_z}{2}+\frac{1}{8}J^{max}$ and $B=-E_z+\frac{1}{4}J^{max}$\\
\hline
\end{tabular}
\end{table}

\bibliographystyle{angewandte}
\bibliography{Ref}
\end{document}